\documentclass[aps,10pt,pre,twocolumn,showpacs]{revtex4-1}
\usepackage{amsmath,amssymb,amsfonts,amsthm}
\usepackage[ascii]{inputenc}
\usepackage{graphicx}
\usepackage{bbm}
\usepackage[caption=false]{subfig}
\usepackage[pdftex,bookmarks=false,colorlinks=true,linkcolor=blue,
citecolor=blue,filecolor=black,urlcolor=blue]{hyperref}

\providecommand{\ud}{\mathrm{d}}

\begin{document}

\title{Thermalization of oscillator chains with onsite anharmonicity and \\ comparison with kinetic theory}

\author{Christian B.~Mendl}
\email{mendl@stanford.edu}
\affiliation{Stanford Institute for Materials and Energy Sciences, SLAC National Accelerator Laboratory and Stanford University, Menlo Park, California 94025, USA}

\author{Jianfeng Lu}
\email{jianfeng@math.duke.edu}
\affiliation{Department of Mathematics, Department of Physics and Department of Chemistry, Duke University, Box 90320, Durham, North Carolina 27708, USA}

\author{Jani Lukkarinen}
\email{jani.lukkarinen@helsinki.fi}
\affiliation{University of Helsinki, Department of Mathematics and Statistics, P.O. Box 68, FI-00014 Helsingin yliopisto, Finland}

\date{\today}

\begin{abstract}
We perform microscopic molecular dynamics simulations of particle chains with an onsite anharmonicity to study relaxation of spatially homogeneous states to equilibrium, and directly compare the simulations with the corresponding Boltzmann--Peierls kinetic theory. The Wigner function serves as common interface between the microscopic and kinetic level. We demonstrate quantitative agreement after an initial transient time interval. In particular, besides energy conservation, we observe the additional quasi-conservation of the phonon density, defined via an ensemble average of the related microscopic field variables and exactly conserved by the kinetic equations. On super-kinetic time scales, density quasi-conservation is lost while energy remains conserved, and we find evidence for eventual relaxation of the density to its canonical ensemble value. However, the precise mechanism remains unknown and is not captured by the Boltzmann--Peierls equations.
\end{abstract}


\maketitle

\section{Introduction}

The arguably most famous system concerning approach to equilibrium on the classical level is the Fermi-Pasta-Ulam (FPU) chain \cite{FPU}. An interesting aspect is the dependence of thermalization on the average energy per degree of freedom \cite{PettiniChaos2005}. For small energies, at least two time scales have been identified, with the short time scale described by a Korteweg-de Vries equation \cite{ZabuskyKruskalPRL1965} (which led to the discovery of solitons), whereas the equipartition of energy modes appears on much longer time scales \cite{BenettinPonno2011}.

In this work, we consider a non-integrable classical oscillator chain with an on-site anharmonicity. From a physics viewpoint, the harmonic chain describes phonons, and the anharmonic term in the Hamiltonian models the main effects of the true interactions between the particles and results in scattering of the phonons. In contrast to the FPU chain, momentum is then not conserved. The oscillator chain has been used to study phonon-mediated thermal conductivity: by regarding the anharmonic term as perturbation of a harmonic crystal, one arrives at the Boltzmann--Peierls equation on the kinetic level, dating back to Peierls in 1929 \cite{Peierls1929}. This kinetic description comprises the thermal conductivity of, e.g., carbon nanotubes \cite{MingoBroido2005}, and was found to nicely agree with microscopic molecular dynamics simulations for such a purpose \cite{DonadioGalli2007, McGaugheyKaviany2004, AokiLukkarinenSpohn2006}. From a mathematical perspective, the phonon Boltzmann--Peierls equation has been justified in \cite{SpohnPhononBoltz2006, LinLepri2016}.

Here, we employ the Boltzmann--Peierls equation as a tool to study thermalization. The Wigner function is the common interface between the microscopic and kinetic level. Defined as ensemble average of the microscopic field variables, the time evolution of this ``microscopic'' Wigner function is expected to solve the kinetic integro-differential equation. Indeed we find quantitative good agreement after an initial transient time interval, see Sec.~\ref{sec:dynamics}. On the kinetic level, the H-theorem states that the entropy must be monotonically increasing, and one can deduce stationary solutions of the Boltzmann--Peierls equation. Thus one is led to the conclusion that the microscopic system has reached thermal equilibrium once the Wigner function has settled to such a stationary kinetic form. However, somewhat reminiscent of prethermalization in quantum systems \cite{SchmiedmayerScience2012}, we find evidence that the kinetic time scale is much shorter than the actual thermalization time scale on the microscopic level. Insight comes from the fact that the kinetic description conserves the phonon density exactly, besides the energy, whereas on the microscopic level only the energy is expected to be conserved. The microscopic phonon density relaxes to its canonical ensemble value only on super-kinetic time scales.

From a broader perspective, our study is particularly motivated by thermalization of isolated quantum systems, which has become an active research field in recent years and still lacks a general understanding \cite{PolkovnikovRev2011}. Cold atom experiments find ``prethermalization'' to a quasi-stationary state described by a generalized Gibbs ensemble, whereas the actual thermalization appears on much longer time scales \cite{SchmiedmayerScience2012}. Even two distinct nonequilibrium steady states have been characterized in the Bose-Hubbard model \cite{KollathPRL2007}, depending on the strength of the Hubbard interaction after a quench, or that a memory of the initial configuration can persist up to the largest simulation times \cite{BanulsCiracHastingsPRL2011}.

\section{Theoretical framework}
\label{sec:framework}

\subsection{Anharmonic chain}

In this work, we consider a classical oscillator chain with an onsite anharmonic potential, as governed by the Hamiltonian
\begin{equation}
\label{eq:H_chain}
H = \sum_{j \in \mathbb{Z}} \Big[ \tfrac{1}{2} p_j^2 + \tfrac{1}{2} \omega_0^2 q_j^2 - \tfrac{1}{2} \delta \omega_0^2 (q_{j-1} q_j + q_j q_{j+1}) + \tfrac{1}{4} \lambda q_j^4 \Big].
\end{equation}
Here $\omega_0^2$ is the strength of the harmonic on-site interaction, $\delta \omega_0^2$ the nearest neighbor interaction, and $\lambda$ the anharmonic on-site interaction, following the notations from \cite{AokiLukkarinenSpohn2006}. For the numerical simulations we will use a system size $L$ such that $j = 0, 1, \dots, L-1$ with periodic boundary conditions.

Fourier transforms are parametrized on their first Brillouin zone chosen as $\mathbbm{T} = [-\frac{1}{2}, \frac{1}{2})$, 
\begin{equation}
\widehat{f}(k) = \sum_{j \in \mathbb{Z}} \mathrm{e}^{-2 \pi i k j} f_j, \quad k \in \mathbbm{T},
\end{equation}
for which the corresponding inverse Fourier transform is
\begin{equation}
f_j = \int_{\mathbbm{T}} \ud k\, \mathrm{e}^{2 \pi i k j} \widehat{f}(k).
\end{equation}
For the simulations we adhere to the same convention; $j \in \mathbbm{Z}$ is replaced by $j = 0, 1, \dots, L-1$ and $\mathbbm{T}$ by a discrete grid with spacing $\frac{1}{L}$, i.e., $k \in \{-\frac{1}{2}, -\frac{1}{2} + \frac{1}{L}, \dots, \frac{1}{2} - \frac{1}{L} \}$.

The dispersion relation of the harmonic part obtained by setting $\lambda=0$ above is given by \cite{AokiLukkarinenSpohn2006, LinLepri2016}
\begin{equation}
\omega(k) = \omega_0 \big(1 - 2 \delta \cos(2\pi k) \big)^{1/2} \, .
\end{equation}
The standard definition of the corresponding phonon mode fields $a(k)$ is then
\begin{equation}
\label{eq:ak_def}
a(k) = \frac{1}{\sqrt{2}} \Big( \sqrt{\omega(k)} \widehat{q}(k) + i \frac{1}{\sqrt{\omega(k)}} \widehat{p}(k) \Big)\, .
\end{equation}
With this choice, if $(q_j(t),p_j(t))$ follows the harmonic evolution, then $a(k,t)$ satisfies $\partial_t a(k,t) = -i \omega(k) a(k,t)$.

We consider here only translation invariant initial data, that is, we suppose that the initial values $(q(0),p(0))$ for the Hamiltonian evolution are chosen from some probability distribution which is invariant under periodic translations of the chain positions. This makes $q(t),p(t)$ random and the corresponding $a(k,t)$ becomes a complex-valued random field. If the initial data of the (anharmonic) evolution is translation invariant and the field has zero mean, $\langle a(k,t) \rangle = 0$, then it is possible to define the Wigner function $W_{\text{mic}}(k, t)$ of the phonon field from the identity
\begin{equation}
\langle a(k,t)^* a(k',t) \rangle = \delta(k - k') W_{\text{mic}}(k, t)\, .
\end{equation}
In this case, the Wigner function can thus be computed via the integral $W_{\text{mic}}(k, t) = \int \ud k'\,\langle a(k,t)^* a(k',t) \rangle$. For the finite periodic lattice used in the numerical simulations, an analogous calculation results in the definition $W_{\text{mic}}(k, t) = L^{-1} \sum_{k'} \langle a(k,t)^* a(k',t) \rangle$. Since only the term with $k'=k$ yields a nonzero value in the sum, we can use the simple expression
\begin{equation}
\label{eq:Wmic_def}
W_{\text{mic}}(k, t) = \frac{1}{L} \langle \lvert a(k,t) \rvert^2 \rangle\,
\end{equation}
to compute the time evolution of the microscopic Wigner function, implicitly also depending on $\lambda$.

\subsection{Kinetic Boltzmann--Peierls equation}

A discussion about the physical and mathematical conditions and prerequisites underlying the phonon Boltzmann equation is given in \cite{SpohnPhononBoltz2006}. Based on these ideas, the phonon Boltzmann equation for the present anharmonic chain has been derived in \cite{AokiLukkarinenSpohn2006}, with further details in \cite{LinLepri2016}.

In accordance with the microscopic model, we consider only spatially homogeneous initial data, for several reasons. First, the inhomogeneities would result in likewise spatially varying temperature distributions. Since the present systems are known to exhibit normal heat conduction \cite{LepriLiviPoliti2003}, the temperature distribution would continue to evolve up to diffusive time scales $\mathcal{O}(L^2)$, where $L$ is the chain length. This additional diffusive evolution complicates the study of thermalization, as it is difficult to separate the various effects. Moreover, for inhomogeneous initial data the anharmonities might lead to corrections at times $\mathcal{O}(\lambda^{-1})$, i.e., before the phonon collisions become active \cite[section 6]{LukkarinenMarcozzi16}, and these corrections should first be taken into account for an honest comparison with kinetic theory.

In the homogeneous case the Wigner function depends only on the phonon wavenumber $k$, as defined above. The associated kinetic equation derived in \cite{AokiLukkarinenSpohn2006,LinLepri2016} reads
\begin{equation}
\label{eq:phonon_boltz_1D}
\partial_t W(k,t) = \mathcal{C}[W(t)](k)
\end{equation}
with the four-phonon collision operator
\begin{equation}
\label{eq:coll1D}
\begin{split}
&\mathcal{C}[W](k_0) = \frac{9\pi}{4} \lambda^2 \int_{\mathbbm{T}^3} \ud k_1 \ud k_2 \ud k_3\, \frac{1}{\omega_0 \omega_1 \omega_2 \omega_3} \\
&\quad \times \delta(\omega_0 + \omega_1 - \omega_2 - \omega_3) \delta(k_0 + k_1 - k_2 - k_3) \\
&\quad \times \big[ W_1 W_2 W_3 + W_0 W_2 W_3 - W_0 W_1 W_3 - W_0 W_1 W_2\big].
\end{split}
\end{equation}
Here we have used the shorthand notation $\omega_j = \omega(k_j)$ and analogously $W_j = W(k_j)$. The time $t$ above has been scaled back to microscopic units which results in the additional factor ``$\lambda^2$'' compared to the formulae in \cite[(3.18)]{AokiLukkarinenSpohn2006}.

The solutions to \eqref{eq:phonon_boltz_1D} preserve the phonon \emph{density},
\begin{equation}
\label{eq:densityWigner}
\rho(t) = \int_{\mathbbm{T}} \ud k\, W(k,t)
\end{equation}
and \emph{energy}
\begin{equation}
\label{eq:energyWigner}
e(t) = \int_{\mathbbm{T}} \ud k\, \omega(k) W(k,t)\,.
\end{equation}
Let us recall that in general the phonon Boltzmann operator includes also collisions which do not preserve the phonon density $\rho(t)$: an example of such a term is obtained by swapping the sign in front of $k_1$ and $\omega_1$ in the conservation delta-functions, as well as the sign of the second term inside the parentheses in (\ref{eq:coll1D}), i.e., by changing the signs from ``$++--$'' to ``$+---$''. (The complete four-phonon collision operator for general dispersion relations is given in \cite[(3.16)]{AokiLukkarinenSpohn2006} and one can check that in general it conserves the energy $e(t)$ but not the density $\rho(t)$.) As proven in \cite[Appendix 18.1]{SpohnPhononBoltz2006}, with more details given in \cite[section 4.2.2]{LinLepri2016}, nearest neighbor dispersion relations do not allow for any (nontrivial) solutions to the two constraints enforced by the delta-functions in the collision operator for any three-phonon collisions, nor for any four-phonon collisions in which the positive and negative contributions do not match each other. For instance, the ``$+---$'' sign-combination leads to a zero contribution to the collision operator, and the remaining terms sum up to \eqref{eq:coll1D}.

\section{Equilibrium Wigner function}
\label{sec:equilWigner}

On the kinetic level, one directly verifies that
\begin{equation}
\label{eq:Weq_kinetic}
W^{\text{eq}}_{\beta',\mu'}(k) = \frac{1}{\beta' (\omega(k) - \mu')}
\end{equation}
are stationary solutions of \eqref{eq:phonon_boltz_1D} and \eqref{eq:coll1D}. As shown in \cite{SpohnPhononBoltz2006,CollisionalInvariants2006}, these are quite generally the only solutions for two- and higher-dimensional crystals. The one-dimensional case is more intricate but we do not expect any new solutions to appear for the nearest neighbor dispersion relations (if $\delta=\frac{1}{2}$, the proof of absence of additional solutions is given in \cite[section 5]{LukkarinenSpohn07}). For an arbitrary Wigner function $W(k)$, the chemical potential $\mu'$ and inverse temperature $\beta'$ of the corresponding thermal equilibrium state on the kinetic level are fixed by the density and energy conservation laws.

\begin{figure}[!ht]
\centering
\subfloat[$\beta = 1$]{
\includegraphics[width=0.475\columnwidth]{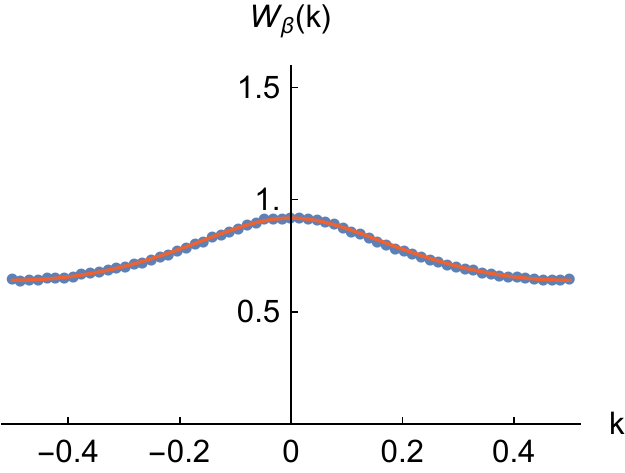}}
\hspace{0.01\columnwidth}
\subfloat[$\beta = 10$]{
\includegraphics[width=0.475\columnwidth]{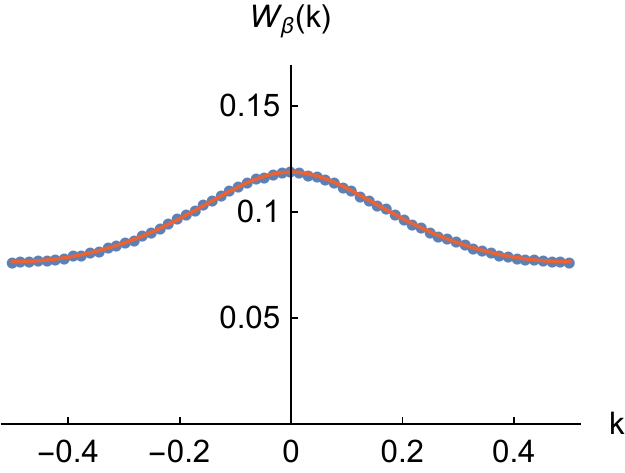}}\\
\subfloat[$\beta = 100$]{
\includegraphics[width=0.475\columnwidth]{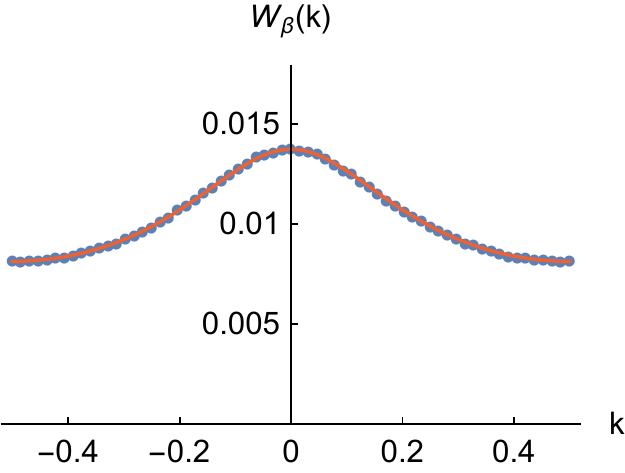}}
\hspace{0.01\columnwidth}
\subfloat[$\beta = 1000$]{
\includegraphics[width=0.475\columnwidth]{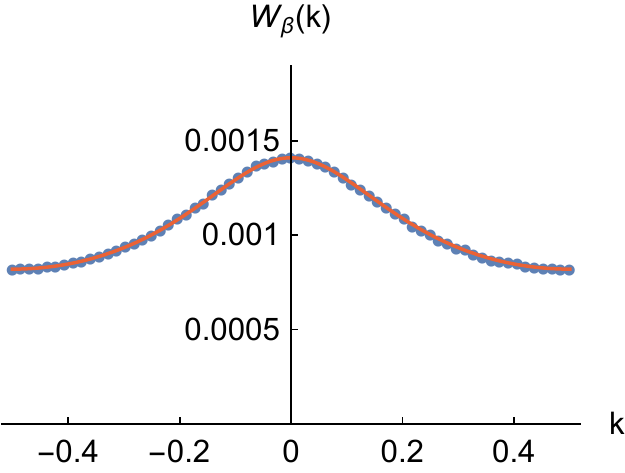}}
\caption{Microscopic thermal Wigner functions $W_{\beta}(k)$ defined in \eqref{eq:Wigner_beta}. The red curves show fitted kinetic equilibrium Wigner functions of the form \eqref{eq:Weq_kinetic}.}
\label{fig:W_thermal}
\end{figure}

A natural question is then whether \eqref{eq:Weq_kinetic} is an accurate description of a \emph{microscopic} thermal Wigner function
\begin{equation}
\label{eq:Wigner_beta}
W_{\beta}(k) = \frac{1}{L} \langle \lvert a(k) \rvert^2 \rangle_{\beta}
\end{equation}
with $a(k)$ defined in \eqref{eq:ak_def} and $\langle \cdot \rangle_{\beta}$ denoting the average with respect to the Gibbs canonical ensemble distribution
\begin{equation}
\label{eq:canonical_ensemble}
Z^{-1}\, \exp[-\beta H].
\end{equation}
To investigate this question numerically, we evaluate \eqref{eq:Wigner_beta} by sampling the microscopic momenta and positions from \eqref{eq:canonical_ensemble}. According to the Hamiltonian \eqref{eq:H_chain}, the momenta $p_j$ are independent Gaussian variables $\sim \exp[-\frac{1}{2} \beta p_j^2]$. The positions decouple from the momenta and are distributed as
\begin{equation}
\label{eq:Gibbs_q}
Z_{\text{sp}}^{-1}\, \exp[{- \beta H_{\text{sp}}}],
\end{equation}
with
\begin{equation}
H_{\text{sp}} = \sum_{j = 0}^{L-1} \Big[ \tfrac{1}{2} \omega_0^2 q_j^2 - \tfrac{1}{2} \delta \omega_0^2 (q_{j-1} q_j + q_j q_{j+1}) + \tfrac{1}{4} \lambda q_j^4 \Big]
\end{equation}
the spatial part of the Hamiltonian (using periodic boundary conditions). To obtain the positions in practice, we discretize the following fictitious overdamped Langevin dynamics \cite{LeimkuhlerMatthews2015} (also known as biased random-walk)
\begin{equation}
\ud q_j(\tau) = - \partial_{q_j} H_{\text{sp}}\,\ud \tau + \sqrt{\tfrac{2}{\beta}}\,\ud W_j(\tau) \,, \quad j = 0,\dots,L-1
\end{equation}
where $W_{j}(\tau)_{\tau \ge 0}$ are standard Wiener processes. Numerically, we use the Euler-Maruyama method with step size $\Delta \tau$,
\begin{equation}
q_j^{n+1} = q_j^n - \Delta\tau\,\partial_{q_j} H_{\text{sp}} + \sqrt{\tfrac{2 \Delta\tau}{\beta}}\,G_j^n \,,
\end{equation}
where the $G_j^n$ are independent standard Gaussian variables. In our implementation, we set $\Delta \tau = \frac{1}{64}$ and perform $1024$ such steps.

The resulting $W_{\beta}(k)$ for four values of $\beta$ is shown in Fig.~\ref{fig:W_thermal} (blue dots). We have set $\omega_0 = 1$, $\delta = \frac{1}{4}$ and $\lambda = 1$ in the microscopic Hamiltonian, with lattice size $L = 64$. By a scaling argument \cite{AokiLukkarinenSpohn2006}, one finds that the anharmonic effects have a strength which depends only on $\lambda/\beta$. Thus for large values of $\beta$ or small $\lambda$, $W_{\beta}(k)$ is expected to be well approximated by $W^{\text{eq}}_{\beta',\mu'}(k)$ in \eqref{eq:Weq_kinetic}. Numerical fits are shown as red curves in Fig.~\ref{fig:W_thermal}; even though $\lambda$ is quite large, one observes very good agreement. The corresponding values of $\beta'$ and $\mu'$ are recorded in the following table:
\begin{center}
\begin{tabular}{r|cccc}
$\beta$  &  1     & 10      & 100       & 1000      \\
\hline
$\beta'$ &  0.912 &  8.98   &  97.1     &  986.4    \\
$\mu'$   & -0.488 & -0.229  &  -0.0426  &   -0.0120
\end{tabular}
\end{center}
One observes that $\beta' \approx \beta$ and $\mu' \approx 0$ for large $\beta$.

\section{Microscopic versus kinetic dynamics and thermalization}
\label{sec:dynamics}

First, we explore at which time scales and with which accuracy the kinetic description is a good approximation to the microscopic dynamics. For that purpose, we compute the time dependent $W_{\text{mic}}(k, t)$ as defined in Eq.~\eqref{eq:Wmic_def} from the microscopic field variables, averaging over many realizations of the microscopic dynamics. This ``microscopic'' Wigner function is then compared to the solution of the kinetic Boltzmann equation in \eqref{eq:phonon_boltz_1D}.

Specifically, we choose the chain length $L = 64$ and Hamiltonian parameters $\omega_0 = 1$, $\delta = 1/4$, $\lambda = 1$ or $\lambda = \frac{1}{2}$. For the numerical time evolution we use the symplectic St\"ormer-Verlet method \cite{HairerLubichWanner2006} with step size $\Delta t = 0.001$. $W_{\text{mic}}(k, t)$ is averaged over $10^5$ realizations of the microscopic dynamics.

\subsection{Bimodal momentum distribution}

Concerning initial states, we sample the initial momenta $p_j$, $j = 0, 1, \dots, L-1$ independently from the bimodal distribution
\begin{equation}
\label{eq:bimod_distr_p_init}
Z_{\text{init}}^{-1}\, \exp\!\left[{- \beta_{\text{init}} \big(4 p_j^4 - \tfrac{1}{2}p_j^2\big)}\right]
\end{equation}
with $\beta_{\text{init}} = 1000$, which by construction markedly differs from the thermodynamic Gaussian equilibrium distribution. On the other hand, the initial positions are sampled from the thermodynamic Gibbs ensemble \eqref{eq:Gibbs_q} with $\beta = \beta_{\text{init}}$. These initial distributions satisfy an important prerequisite for the theoretical derivation of the kinetic Boltzmann equation, namely translational invariance. We set $\lambda = 1$ in what follows.

\begin{figure}[!ht]
\centering
\subfloat[$t = 0$]{
\includegraphics[width=0.475\columnwidth]{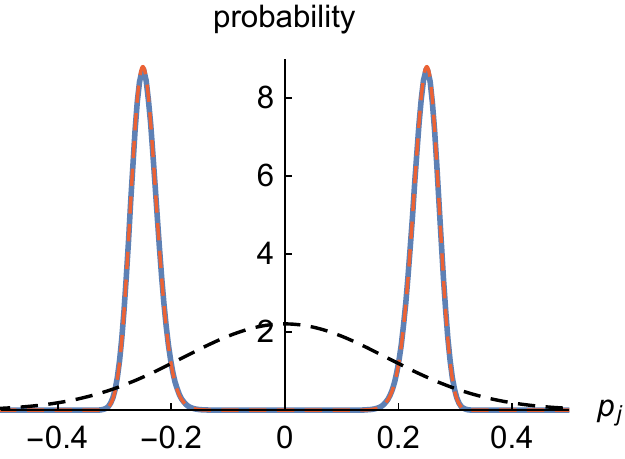}}
\hspace{0.01\columnwidth}
\subfloat[$t = 4$]{
\includegraphics[width=0.475\columnwidth]{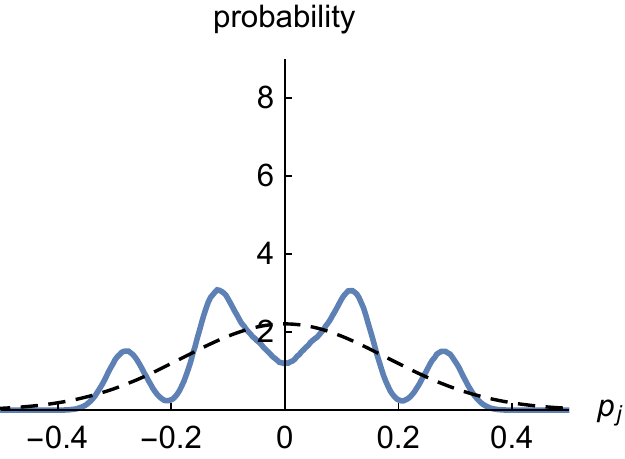}}\\
\subfloat[$t = 8$]{
\includegraphics[width=0.475\columnwidth]{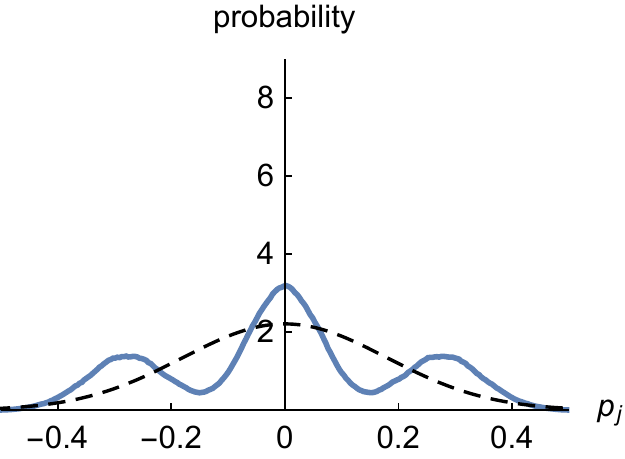}}
\hspace{0.01\columnwidth}
\subfloat[$t = 16$]{
\includegraphics[width=0.475\columnwidth]{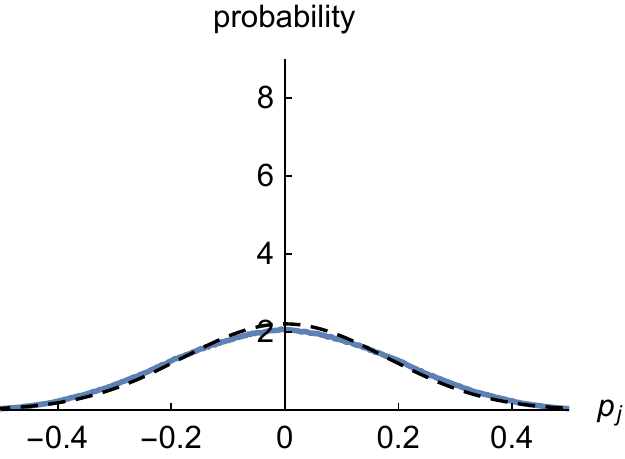}}\\
\subfloat[$t = 150$]{
\includegraphics[width=0.475\columnwidth]{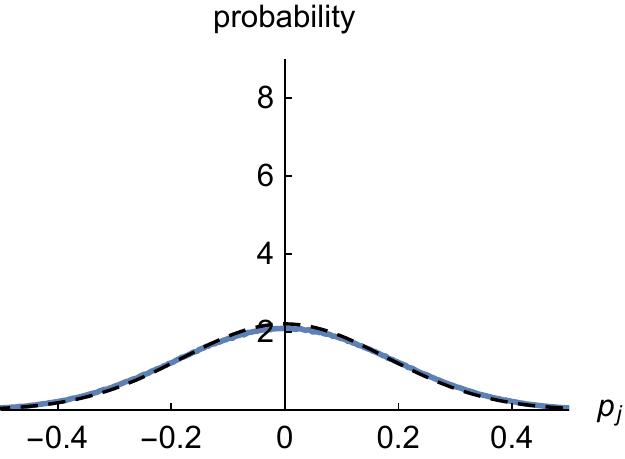}}
\hspace{0.01\columnwidth}
\subfloat[$t = 500$]{
\includegraphics[width=0.475\columnwidth]{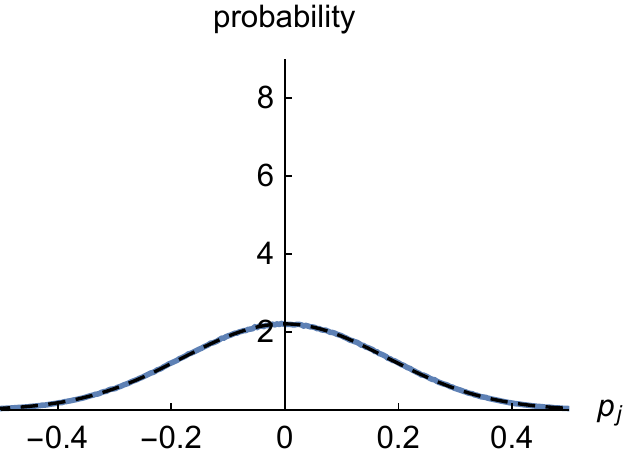}}
\caption{Time evolution of the momentum distribution (blue curve), based on the microscopic dynamics and initialized via \eqref{eq:bimod_distr_p_init} (red dashed) at $t = 0$.}
\label{fig:bimod_hist_p_evolution}
\end{figure}

\begin{figure}[!ht]
\centering
\includegraphics[width=0.7\columnwidth]{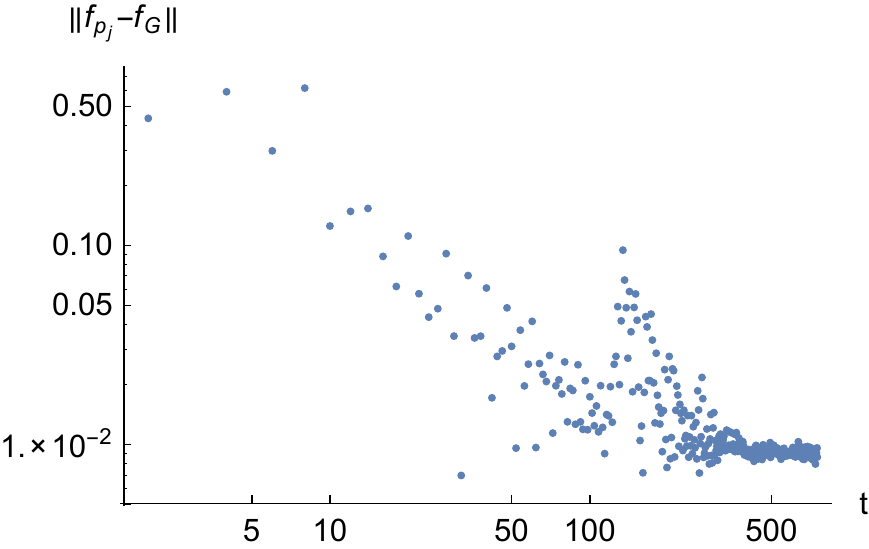}
\caption{Convergence of the momentum distribution shown in Fig.~\ref{fig:bimod_hist_p_evolution} towards the stationary Gaussian distribution \eqref{eq:gauss_p_thermal}.}
\label{fig:bimod_hist_p_convergence}
\end{figure}

As preliminary characterization of the microscopic dynamics, we compute histograms of the momenta $p_j(t)$ to obtain the time evolution of the momentum distribution. It is shown as blue curve in Fig.~\ref{fig:bimod_hist_p_evolution} for several time points. The red dashed curve at $t = 0$ is the initial distribution \eqref{eq:bimod_distr_p_init}, which -- by construction -- precisely matches the numerical distribution. One observes that the momentum distribution relaxes very quickly to a Gaussian-shaped curve already at $t = 16$. Nevertheless, it keeps slightly oscillating until finally settling towards a Gaussian distribution
\begin{equation}
\label{eq:gauss_p_thermal}
\frac{1}{\sqrt{2\pi/\beta_{\text{th}}}} \exp\!\left[{- \tfrac{1}{2} \beta_{\text{th}}\,p^2}\right]
\end{equation}
with fitted $\beta_{\text{th}} = 30.74$, shown as black dashed curve in Fig.~\ref{fig:bimod_hist_p_evolution}. The $L^2$-norm distance to this Gaussian distribution is quantified in Fig.~\ref{fig:bimod_hist_p_convergence}. Following the initial exponential convergence up to around $t = 50$, one observes a temporary increase of the distance between $t = 100$ and $t = 200$, corresponding to the mentioned slight oscillations. At around $t = 500$ the momentum distribution has reached stationarity.

\begin{figure}[!ht]
\centering
\subfloat[$t = 0$]{
\includegraphics[width=0.475\columnwidth]{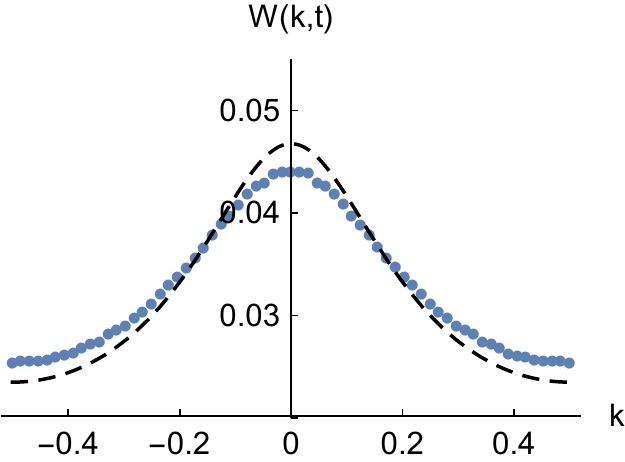}}
\hspace{0.01\columnwidth}
\subfloat[$t = 150$]{
\includegraphics[width=0.475\columnwidth]{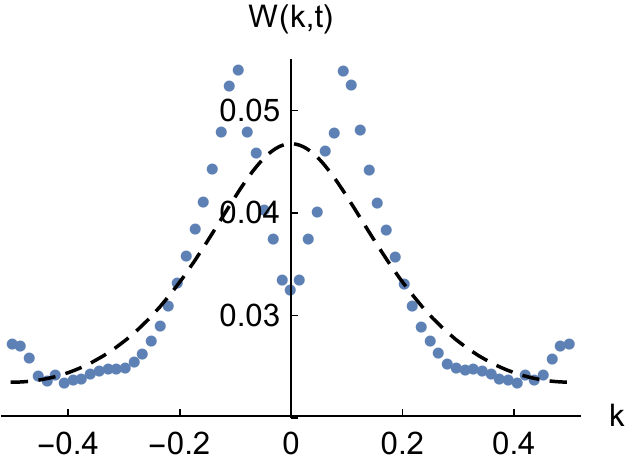}}\\
\subfloat[$t = 250$]{
\includegraphics[width=0.475\columnwidth]{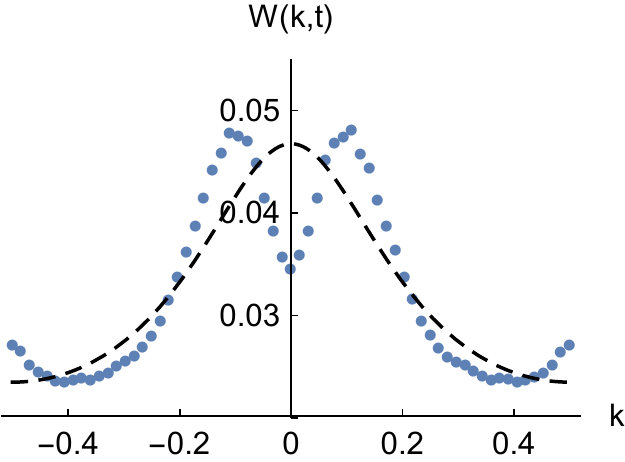}}
\hspace{0.01\columnwidth}
\subfloat[$t = 500$]{
\includegraphics[width=0.475\columnwidth]{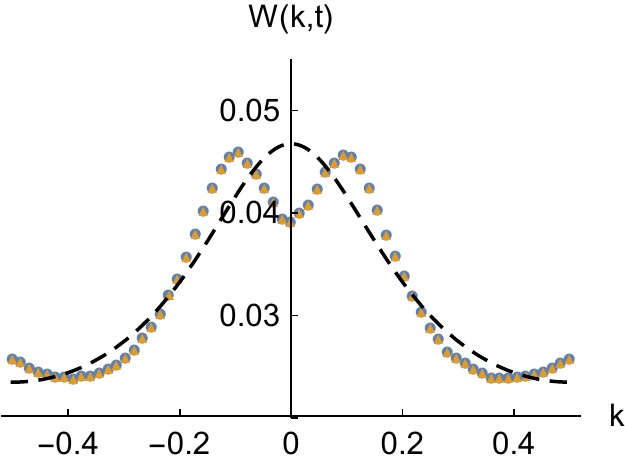}}\\
\subfloat[$t = 1000$]{
\includegraphics[width=0.475\columnwidth]{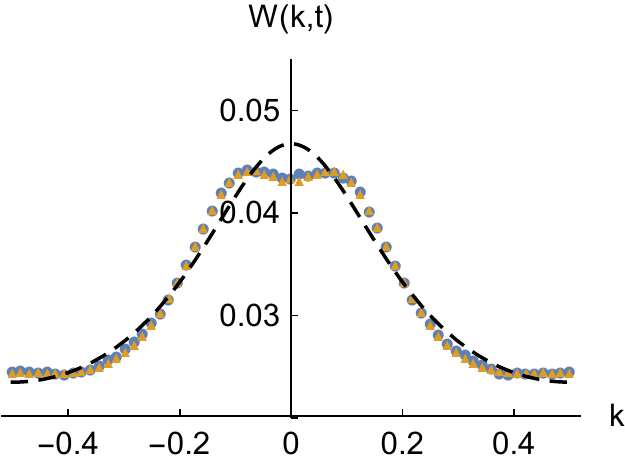}}
\hspace{0.01\columnwidth}
\subfloat[$t = 2500$]{
\includegraphics[width=0.475\columnwidth]{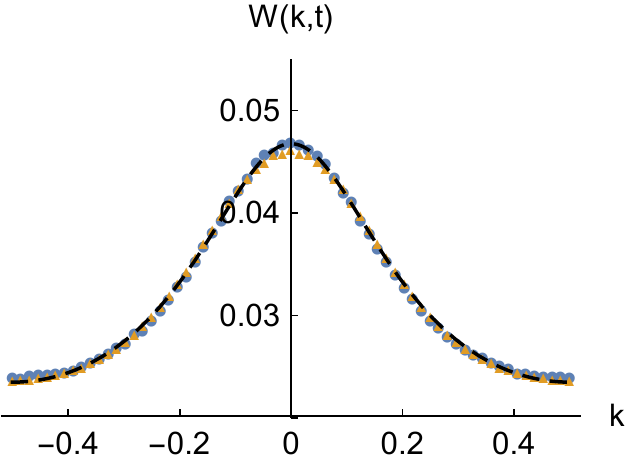}}
\caption{Time evolution of the Wigner function computed from microscopic field variables (blue dots), in comparison with a simulation of the kinetic Boltzmann equation (yellow triangles) starting at $t = 500$.}
\label{fig:bimod_W_time_evolution}
\end{figure}

Based on the microscopic field variables, we now compute the time evolution of the Wigner function via \eqref{eq:Wmic_def}, which is shown as blue dots in Fig.~\ref{fig:bimod_W_time_evolution}. Around $t = 150$ (in the oscillatory regime of the momentum distribution), it exhibits a somewhat ``excited'' form. Surprisingly, at $t = 500$ when the momentum distribution has already reached stationarity, the Wigner function clearly differs from an equilibrium shape as discussed in Sec.~\ref{sec:equilWigner}. On the other hand, from around $t = 500$ one observes that the microscopic Wigner function starts a smooth transition towards the black dashed curve $1/(\beta' (\omega(k) - \mu'))$ in Fig.~\ref{fig:bimod_W_time_evolution} with fitted $\beta' = 40.91$ and $\mu' = 0.1842$. One is drawn to hypothesize that the kinetic description is applicable in this regime. To quantify, we numerically solve the kinetic Boltzmann equation \eqref{eq:phonon_boltz_1D} starting at $t = 500$, with the initial kinetic Wigner function set equal to the microscopic Wigner function at $t = 500$. For the kinetic collision operator we adapt the numerical procedure described in \cite{BoltzmannNonintegrable2013} to the present setting. The kinetic solution is sped up by a factor $4/3$ to align it with the microscopic time evolution, and shown superimposed as yellow triangles in Fig.~\ref{fig:bimod_W_time_evolution}. The microscopic and kinetic Wigner functions agree fairly well. The required $4/3$ time scaling factor could stem from the relatively large anharmonic coefficient $\lambda = 1$; namely, the theoretical derivation of the kinetic Boltzmann equation uses an expansion for small $\lambda$. To check, we have repeated the simulation with the same parameters except $\lambda = \frac{1}{2}$, with qualitatively very similar results (data not shown). Besides the theoretical time scale $\lambda^{-2} t$, the correction speed up factor is now approximately $1/0.85$, i.e., closer to $1$ than for $\lambda = 1$.

\begin{figure}[!ht]
\centering
\includegraphics[width=0.7\columnwidth]{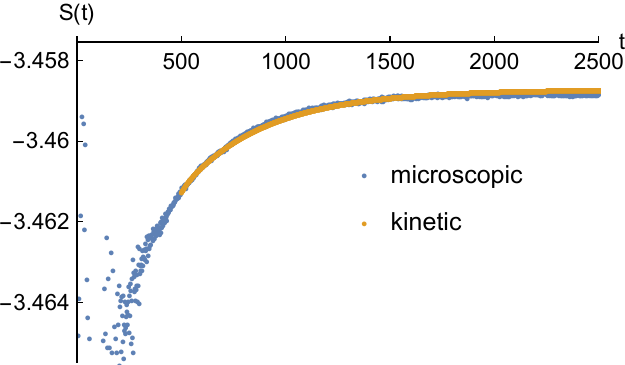}
\caption{Time evolution of the entropy computed via \eqref{eq:entropyWigner}, superimposing the microscopic with the kinetic simulation.}
\label{fig:bimod_entropy_evolution}
\end{figure}

\begin{figure}[!b]
\centering
\subfloat[$\rho_{\text{mic}}(t) - \rho_{\text{mic}}(t_{\max})$]{
\includegraphics[width=0.5\columnwidth]{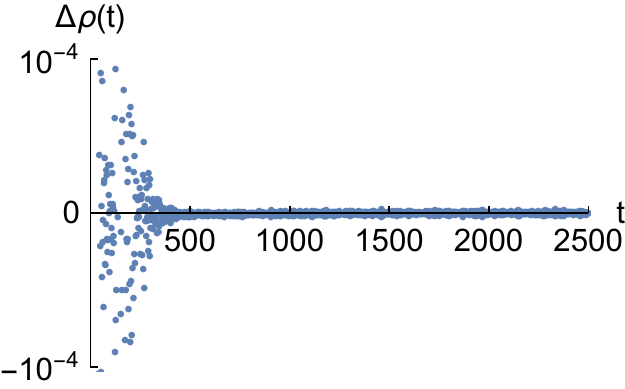}}
\subfloat[$e_{\text{mic}}(t) - e_{\text{mic}}(t_{\max})$]{
\includegraphics[width=0.5\columnwidth]{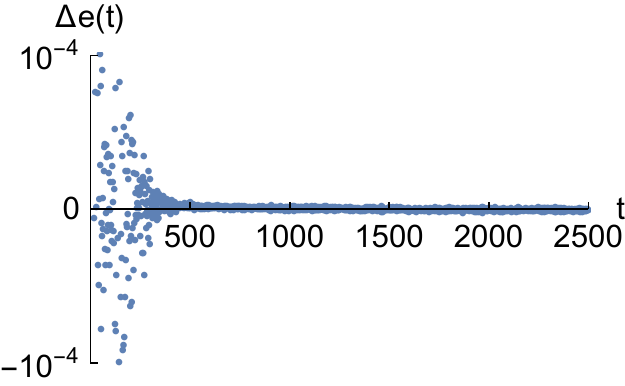}}
\caption{Time evolution of the density and energy difference computed via \eqref{eq:densityWigner} and \eqref{eq:energyWigner} from the microscopic Wigner function shown in Fig.~\ref{fig:bimod_W_time_evolution}.}
\label{fig:bimod_density_energy_mic}
\end{figure}

The entropy based on the phonon Wigner function is defined as
\begin{equation}
\label{eq:entropyWigner}
S(t) = \int_{\mathbbm{T}} \ud k\,\log W(k,t).
\end{equation}
On the kinetic level, the H-theorem states that the entropy is monotonically increasing with time. Fig.~\ref{fig:bimod_entropy_evolution} compares the entropy computed from the microscopic Wigner function with the entropy based on the Wigner function on the kinetic level, starting at $t = 500$ and using the same $4/3$ time scaling factor as before. As expected from Fig.~\ref{fig:bimod_W_time_evolution}, the agreement is fairly good, and the entropy is indeed monotonically increasing.

Finally, Fig.~\ref{fig:bimod_density_energy_mic} shows the time evolution of the density and energy computed via \eqref{eq:densityWigner} and \eqref{eq:energyWigner} from the microscopic Wigner function, after subtracting the final density and energy, respectively. Their numerical values are $\rho_{\text{mic}}(t_{\max}) = 0.0324$ and $e_{\text{mic}}(t_{\max}) = 0.0304$. One can discriminate an initial time interval up to $t \approx 500$ where the density and energy are not exactly preserved. After that, density and energy conservation holds quite precisely, in accordance with the kinetic Boltzmann equation. Nevertheless, the density conservation is surprising since the Wigner function is computed from the microscopic fields, but on the microscopic level only the energy is expected to be conserved.

\subsection{Random phase approximation for the initial Wigner function}

In the previous subsection we have specified the microscopic Wigner function at $t = 0$ only indirectly via the position and momentum distributions. An alternative route to start from an arbitrary prescribed Wigner function $W_0(k)$ consists of ``inverting'' Eq.~\eqref{eq:Wmic_def} and setting
\begin{equation}
\label{eq:ak_random_phase}
a(k) = \sqrt{L\,W_0(k)}\,\mathrm{e}^{i \varphi(k)},
\end{equation}
with uniformly distributed random variables $\varphi(k) \in [0, 2\pi]$, independently for each discretized $k$. One can check that all odd moments of the field $a$ are then initially zero, since $\langle \mathrm{e}^{i \varphi(k)} \rangle = 0$, $\langle \mathrm{e}^{i (\varphi(k) + \varphi(k'))} \rangle = 0$, etc. It straightforwardly follows that the odd moments are zero for all times $t$, and we find in particular that $\langle a(k,t)\rangle=0$. By these properties the initial, hence also time-evolved, field is translation invariant: lattice translation by $x_0$ corresponds to multiplication of $a(k)$ by a factor $\mathrm{e}^{2\pi i k x_0}$, i.e., a nonrandom phase. As $\varphi(k)$ is uniformly distributed on $[0,2\pi]$, the random variables $\mathrm{e}^{i \varphi(k)}$ and $\mathrm{e}^{i (\varphi(k)+2\pi k x_0)}$ have the same distribution for all $k$, $x_0$.

In the numerical simulation we set
\begin{equation}
\label{eq:W0_random_phase}
W_0(k) = \tfrac{1}{40} \Big(\sin(2 \pi k)^2 + \mathrm{e}^{\sin(\pi (k - 1/3))^2 - 1} + \tfrac{1}{4} \cos(8 \pi k)\Big).
\end{equation}
For each realization of $a(k)$, the initial microscopic positions and momenta are determined by inverting Eq.~\eqref{eq:ak_def}. We then solve the microscopic time evolution as before, with parameters $L = 64$, $\omega_0 = 1$, $\delta = 1/4$ and $\lambda = \frac{1}{2}$. The time dependent momentum and position distribution and microscopic Wigner function is averaged over $10^5$ such realizations.

\begin{figure}[!ht]
\centering
\subfloat{
\label{fig:random_phase_hist_p_time}
\includegraphics[width=0.475\columnwidth]{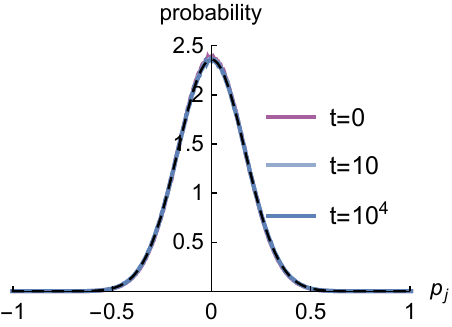}}
\hspace{0.01\columnwidth}
\subfloat{
\includegraphics[width=0.475\columnwidth]{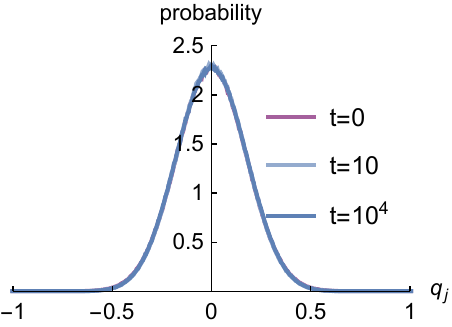}}
\caption{Momentum and position distributions based on the microscopic dynamics with initial ``random phase'' field \eqref{eq:ak_random_phase}, staying constant in time. The black dashed line on the left is a Gaussian distribution \eqref{eq:gauss_p_thermal} with fitted $\beta_{\text{th}} = 35.00$.}
\label{fig:random_phase_hist_pq_time}
\end{figure}

Different from the previous simulation example, one now observes that the momentum and position distributions remain constant in time, as shown in Fig.~\ref{fig:random_phase_hist_pq_time} with three time points superimposed. Analogous to the previous analysis, the black dashed line in the left subfigure is a Gaussian distribution \eqref{eq:gauss_p_thermal} with fitted $\beta_{\text{th}} = 35.00$.

\begin{figure}[!ht]
\centering
\subfloat[$t = 0$]{
\includegraphics[width=0.475\columnwidth]{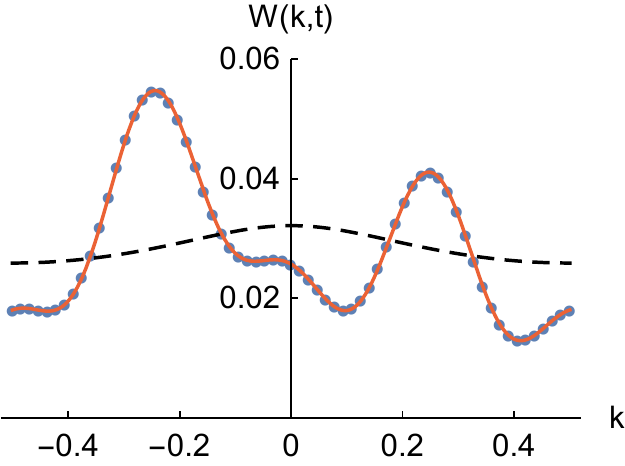}}
\hspace{0.01\columnwidth}
\subfloat[$t = 250$]{
\label{fig:random_phase_W_t250}
\includegraphics[width=0.475\columnwidth]{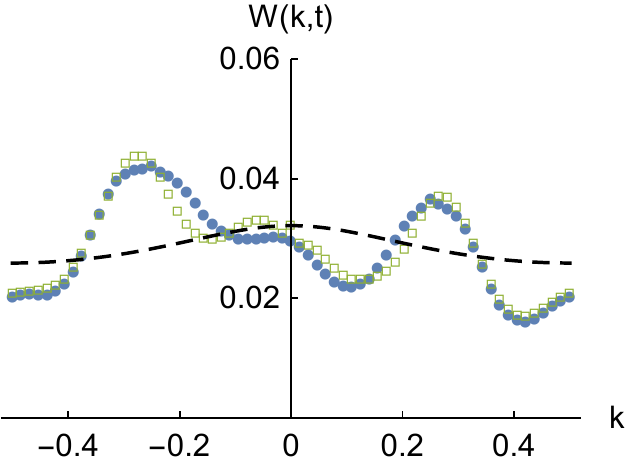}}\\
\subfloat[$t = 500$]{
\includegraphics[width=0.475\columnwidth]{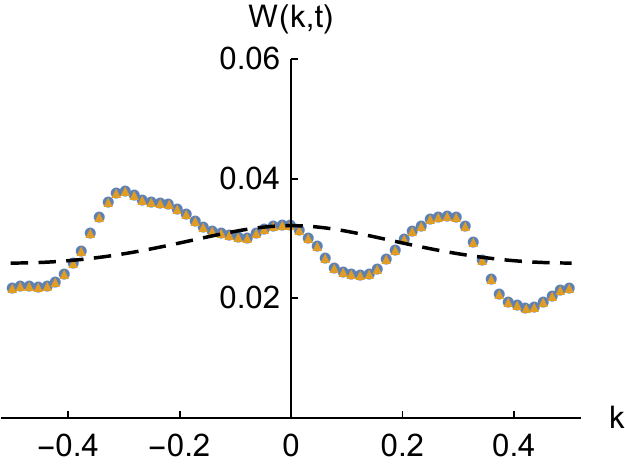}}
\hspace{0.01\columnwidth}
\subfloat[$t = 1000$]{
\includegraphics[width=0.475\columnwidth]{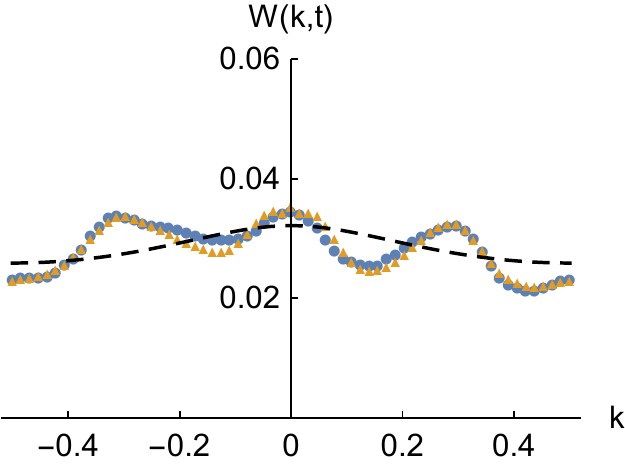}}\\
\subfloat[$t = 5000$]{
\includegraphics[width=0.475\columnwidth]{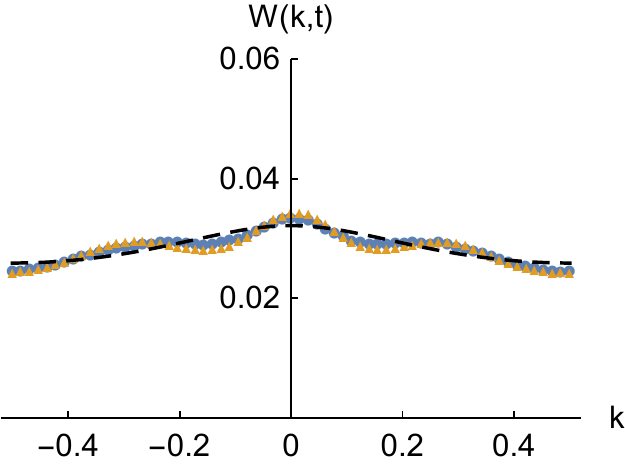}}
\hspace{0.01\columnwidth}
\subfloat[$t = 10000$]{
\label{fig:random_phase_W_tmax}
\includegraphics[width=0.475\columnwidth]{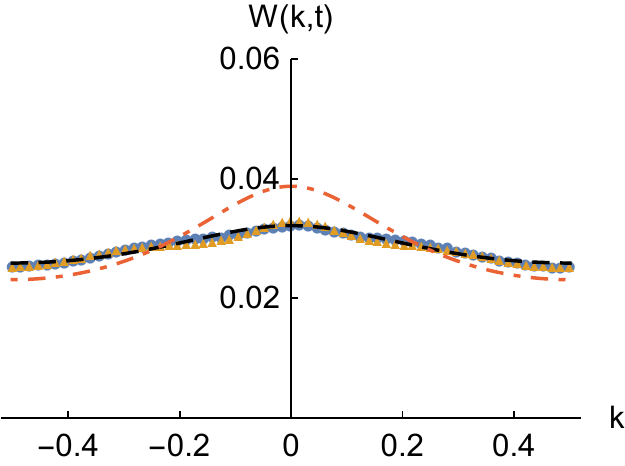}}
\caption{Time evolution of the Wigner function computed from microscopic field variables (blue dots) with initial ``random phase'' field \eqref{eq:ak_random_phase}, in comparison with the kinetic Boltzmann solution (yellow triangles) starting at $t = 500$.}
\label{fig:random_phase_W_time_evolution}
\end{figure}

The time evolution of the microscopic Wigner function is visualized as blue dots in Fig.~\ref{fig:random_phase_W_time_evolution}. The faint red curve at $t = 0$ shows the analytic form in \eqref{eq:W0_random_phase}. Note that the equilibration time is now longer, as expected from the theoretical $\lambda^{-2} t$ time scale with $\lambda = \frac{1}{2}$ instead of $\lambda = 1$. As before, there seems to be an initial transient time period which is not well described by the kinetic Boltzmann equation. This is noticeable when comparing the microscopic Wigner function with the kinetic solution starting at $t = 0$, shown as superimposed green squares at $t = 250$ in Fig.~\ref{fig:random_phase_W_t250}. For the present simulation, the extent of this transient period is not discernible from the momentum distribution, and currently we are not aware of a good indicator. Nevertheless, one expects that the kinetic Boltzmann description becomes applicable after this transient period. As test, we compare the microscopic Wigner function with the kinetic solution starting at $t = 500$, shown as yellow triangles in Fig.~\ref{fig:random_phase_W_time_evolution}. The time evolution is adjusted by the theoretical $\lambda^{-2} t$ scale, without any further correction factor. The agreement is reasonably good, but not as precise as the simulation results in Fig.~\ref{fig:bimod_W_time_evolution}. The deviation could stem from the fact that the transient period extends beyond $t = 500$, or the still relatively large $\lambda = \frac{1}{2}$ compared to the theoretical perturbation expansion for small $\lambda$.

\begin{figure}[!ht]
\centering
\includegraphics[width=0.7\columnwidth]{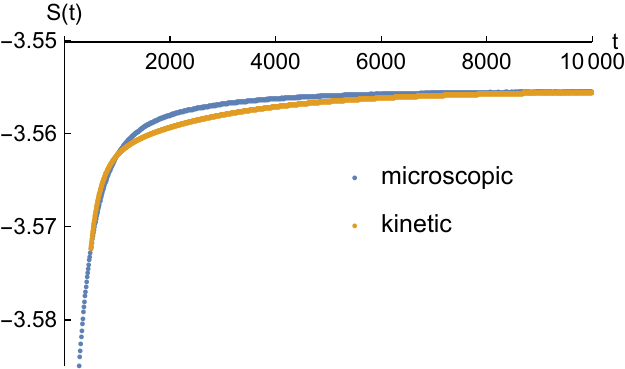}
\caption{Time evolution of the entropy based on the Wigner functions in Fig.~\ref{fig:random_phase_W_time_evolution}, superimposing the microscopic ``random phase'' simulation with the kinetic solution starting at $t = 500$.}
\label{fig:random_phase_entropy_evolution}
\end{figure}

Fig.~\ref{fig:random_phase_entropy_evolution} shows a comparison of the microscopic and kinetic entropies. As expected from Fig.~\ref{fig:random_phase_W_time_evolution}, the agreement is less precise than in Fig.~\ref{fig:bimod_entropy_evolution}; nevertheless, the values at the largest time are very close, and the entropies are monotonically increasing, in accordance with the H-theorem.

\begin{figure}[!ht]
\centering
\subfloat[$\rho_{\text{mic}}(t) - \rho_{\text{mic}}(t_{\max})$]{
\label{fig:random_phase_density_mic}
\includegraphics[width=0.5\columnwidth]{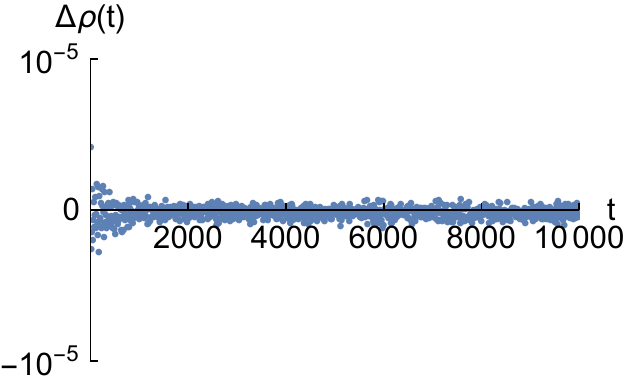}}
\subfloat[$e_{\text{mic}}(t) - e_{\text{mic}}(t_{\max})$]{
\includegraphics[width=0.5\columnwidth]{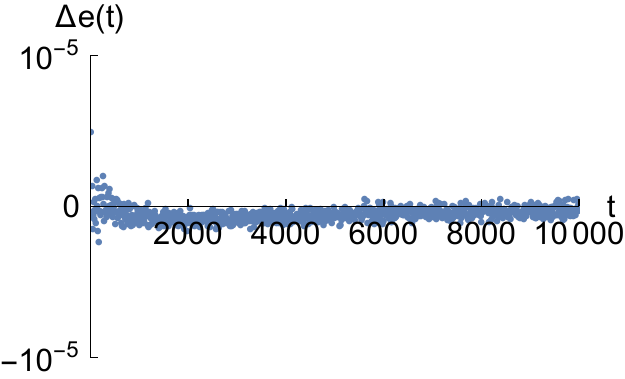}}
\caption{Time evolution of the density and energy difference based on the microscopic Wigner function shown in Fig.~\ref{fig:random_phase_W_time_evolution}.}
\label{fig:random_phase_density_energy_mic}
\end{figure}

The time evolution of the density and energy based on the microscopic Wigner function is visualized in Fig.~\ref{fig:random_phase_density_energy_mic}, after subtracting the density $\rho_{\text{mic}}(t_{\max}) = 0.0287$ and energy $e_{\text{mic}}(t_{\max}) = 0.0278$ at the largest simulation time $t_{\max} = 10000$, respectively. One observes that both are conserved very well during the time evolution, with deviations on the order of $10^{-6}$, which are presumably due to statistical noise. We use the density and energy at $t_{\max}$ to compute the kinetic equilibrium Wigner function of the form $1/(\beta' (\omega(k) - \mu'))$, see Eq.~\eqref{eq:Weq_kinetic}. That is, we fit $\beta'$ and $\mu'$ numerically such that $1/(\beta' (\omega(k) - \mu'))$ has density $\rho_{\text{mic}}(t_{\max})$ and energy $e_{\text{mic}}(t_{\max})$, obtaining $\beta' = 14.50$ and $\mu' = -1.438$. The result is shown as black dashed curve in Fig.~\ref{fig:random_phase_W_time_evolution}, which fits the microscopic Wigner function at the largest simulation time very well.

On the other hand, we have seen in Fig.~\ref{fig:random_phase_hist_p_time} that the ``microscopic'' inverse temperature equals $\beta_{\text{th}} = 35$. Repeating the computations in Sec.~\ref{sec:equilWigner} for $\beta_{\text{th}}$, one obtains the microscopic thermal Wigner function $W_{\beta_{\text{th}}}(k)$ shown as dot-dashed red curve in Fig.~\ref{fig:random_phase_W_tmax}. Surprisingly, $W_{\beta_{\text{th}}}(k)$ clearly deviates from the microscopic Wigner function at $t_{\max}$. Our interpretation is that the microscopic Wigner function has not reached its asymptotic $t \to \infty$ form, but is constrained by the quasi-conserved density shown in Fig.~\ref{fig:random_phase_density_mic}. The Wigner function should eventually converge to $W_{\beta_{\text{th}}}(k)$ for much longer simulation times, since only the energy is conserved on the microscopic level. As an appreciation of the long time scales, the entropy of $W_{\beta_{\text{th}}}(k)$ equals $-3.54275$, which is noticeably larger than the entropies shown in Fig.~\ref{fig:random_phase_entropy_evolution}.

\begin{figure}[!ht]
\centering
\subfloat[$\rho_{\text{mic}}(t) - \rho_{\text{mic}}(t_{\max})$]{
\includegraphics[width=0.5\columnwidth]{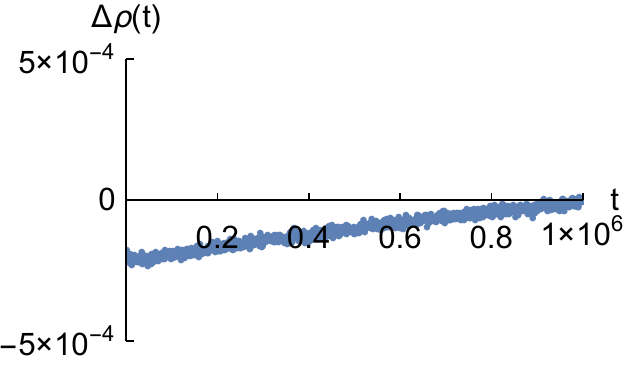}}
\subfloat[$e_{\text{mic}}(t) - e_{\text{mic}}(t_{\max})$]{
\includegraphics[width=0.5\columnwidth]{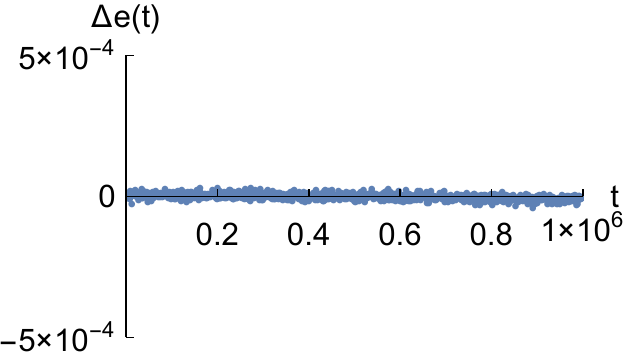}}
\caption{Time evolution of the density and energy difference using the same simulation parameters as in Fig.~\ref{fig:random_phase_density_energy_mic}, except for larger $\lambda = 10$ and longer simulation time $t_{\max} = 10^6$. Note the different scale of the $y$-axis compared to Fig.~\ref{fig:random_phase_density_energy_mic}.}
\label{fig:random_phase_lambda10_density_energy_mic}
\end{figure}

To check the long-time convergence hypothesis, we have repeated the simulation with the same initial Wigner function and simulation parameters, except for larger $\lambda = 10$ to increase the anharmonic effects, and longer simulation time $t_{\max} = 10^6$. Fig.~\ref{fig:random_phase_lambda10_density_energy_mic} shows the time evolution of the density and energy analogous to Fig.~\ref{fig:random_phase_density_energy_mic}. Indeed one observes a drift of the density, while the energy remains constant (up to statistical noise). Due to the increased computational effort associated with $t_{\max} = 10^6$, averages are taken with respect to $10^4$ realizations instead of $10^5$. Even at $t_{\max}$, the microscopic Wigner function still deviates from the microscopic thermal Wigner function (data not shown), but the drift of the density and conservation of the energy supports the expected convergence at even longer time scales.

\section{Conclusions}

The kinetic Boltzmann--Peierls description cannot account for the eventual density relaxation, but nevertheless explains one time scale of the equilibration process. Moreover, it is quite insensitive to the anharmonic strength $\lambda$: in our simulations with relatively large $\lambda = \frac{1}{2}$ and $\lambda = 1$, good agreement between the microscopic and kinetic Wigner functions is found, even for much longer than suggested by the $\mathcal{O}(\lambda^{-2})$ kinetic scaling limit. While this can be partially explained by the small values of the initial fields, i.e., by focusing on low temperatures which subdue the anharmonic effects, the agreement for large $\lambda$ is still surprising.

A natural follow-up question is whether kinetic descriptions can be likewise utilized for understanding thermalization of quantum systems, and whether such descriptions are applicable beyond weak interactions. While the microscopic simulation of interacting quantum systems is usually much more demanding \cite{MartinReiningCeperley2016} than classical molecular dynamics simulations, there is little difference in solving their respective kinetic limits, the quantum Boltzmann--Nordheim and Boltzmann--Peierls equations. The former traces back to Nordheim in 1928 \cite{Nordheim1928} and is also referred to as Uehling-Uhlenbeck equation \cite{UehlingUhlenbeck1933}. A derivation without spin can be found in \cite{LukkarinenSpohn2009}, and an analogous derivation and simulation for  the standard fermionic Hubbard-model with spin is provided in \cite{FurstLukkarinenSpohnMei2013, LukkarinenSpohnMei2015, FurstMendlSpohn2012}. Similar kinetic descriptions have indeed already been used to study thermalization in the Hubbard model \cite{MoeckelKehreinPRL2008, StarkKollar2013}, but much is still left as future work.

\textbf{Acknowledgments.} We would like to thank Herbert Spohn for many helpful discussions. C.~B.~Mendl acknowledges the hospitality of Duke university during a research visit, and support from the Humboldt foundation via a Feodor Lynen fellowship. The work of J.~Lu is supported in part by the National Science Foundation under grant DMS-1454939. J.~Lukkarinen has been supported by the Academy of Finland via the Centre of Excellence in Analysis and Dynamics Research (project 271983) and from an Academy Project (project 258302). The numerical simulations for this work used computational resources of the Leibniz-Rechenzentrum, M\"unchen.


\end{document}